\documentclass[12pt]{article}
\usepackage{graphicx}
\RequirePackage{xspace}

\newcommand\pubdate{September 7, 2011}

\newcommand\pubnumber{arXiv:1109.1343}

\def\Title#1{\begin{center} {\Large #1 } \end{center}}
\def\Author#1{\begin{center}{ \sc #1} \end{center}}
\def\Address#1{\begin{center}{ \it #1} \end{center}}

\newcommand\pubblock{\rightline{\begin{tabular}{l} \pubnumber\\
         \pubdate  \end{tabular}}}
\newenvironment{Abstract}{\begin{center}{\bf Abstract}\end{center} \bigskip \begin{quotation}  }{\end{quotation}}
\newenvironment{Presented}{\begin{quotation} \begin{center} 
             PRESENTED AT\end{center}\bigskip 
      \begin{center}\begin{large}}{\end{large}\end{center} \end{quotation}}
\def\Acknowledgements{\bigskip  \bigskip \begin{center} \begin{large}
             \bf ACKNOWLEDGEMENTS \end{large}\end{center}}

\def\beq{\begin{equation}}
\def\eeq#1{\label{#1}\end{equation}}
\def\eeqn{\end{equation}}

\def\beqa{\begin{eqnarray}}
\def\eeqa#1{\label{#1}\end{eqnarray}}
\def\eeqan{\end{eqnarray}}

\let\bar=\overbar

\def\Dslash{\not{\hbox{\kern-4pt $D$}}}
\def\dslash{\not{\hbox{\kern-2pt $\del$}}}

\def\msb{{\bar{\ssstyle M \kern -1pt S}}}

\def\CP      {\ensuremath{C\!P}\xspace}
\def\Abar    {\kern 0.18em\overline{\kern -0.18em{\cal A}}{}\xspace}
\def\Bbar    {\kern 0.18em\overline{\kern -0.18em B}{}\xspace}
\mathchardef\Upsilon="7107
\def\Y#1S{\ensuremath{\Upsilon{(#1S)}}\xspace}
\def\FourS {\Y4S}

\begin{document}
\begin{titlepage}
\pubblock

\vfill


\Title{Measurement of the UT angle $\phi_2$}
\vfill
\Author{Gagan B. Mohanty}  
\Address{Tata Institute of Fundamental Research\\ Homi Bhabha Road, Mumbai 400 005, INDIA}
\vfill


\begin{Abstract}
We give a status report on measurements of the angle $\phi_2(\alpha)$ of
the CKM unitarity triangle (UT) and the so-called $K\pi$ puzzle. Results
presented are mostly from the two $B$-factory experiments, Belle and BaBar.
\end{Abstract}

\vfill

\begin{Presented}
The Ninth International Conference on\\
Flavor Physics and CP Violation\\
(FPCP 2011)\\
Maale Hachamisha, Israel,  May 23--27, 2011
\end{Presented}
\vfill

\end{titlepage}
\def\thefootnote{\fnsymbol{footnote}}
\setcounter{footnote}{0}


\section{Introduction}

$B$ meson decays provide an excellent laboratory to test
$\CP$ violation mechanism in the standard model (SM),
which is attributed to an irreducible phase appearing in the
quark-flavor mixing matrix, known as the Cabibbo-Kobayashi-Maskawa
(CKM) matrix~\cite{ref:ckm}. Unitarity of the $3\times 3$
CKM matrix can be conveniently depicted as a set of triangles
in the complex plane, one of which is the so-called unitarity
triangle (UT) that connects matrix elements of the first and
third row. One of the major goals for heavy flavor experiments
of the last few decades -- be it the two $B$-factory experiments
of Belle~\cite{ref:belle} and BaBar~\cite{ref:babar}, or the
recently-started LHCb experiment~\cite{ref:lhcb}
-- has been to precisely measure the sides and angles ($\phi_1$,
$\phi_2$ and $\phi_3$)\footnote{An equally popular notation of
$\alpha$, $\beta$ and $\gamma$ is also available in the literature.}
of the UT. In doing so, experimenters aim to verify whether the
CKM framework is the correct description of $\CP$ violation in
the SM, and to constrain possible new physics effects that could
lead to internal inconsistencies among various measurements.
In these proceedings, we present a minireview on the measurement
of the angle $\phi_2$ carried out at Belle and BaBar, followed by
a brief report on the $K\pi$ puzzle.

\section{Enter the angle {\boldmath $\phi_2$}}

The UT angle $\phi_2={\rm arg}(-V_{td}V^*_{tb}/V_{ud}V^*_{ub})$
brings to fore two CKM matrix elements that are complex at
the lowest order: $V_{ub}$, involved in decays proceeding via
a $b\to u$ tree-level transition with phase $-\phi_3$, and
$V_{td}$, involved in $B^0\Bbar^0$ mixing having phase equals
to $-\phi_1$. Therefore, $\phi_2=\pi-\phi_1-\phi_3$ can be
extracted from time-dependent $\CP$ violation asymmetries in
the interference between decay and mixing in $b\to u$
tree-dominated decays of neutral $B$ mesons, such as
$B^0\to\pi^+\pi^-$, $\rho^+\rho^-$ and $\pi^+\pi^-\pi^0$.

At the $B$ factories, the measured time-dependent $\CP$ asymmetry
can be given by
\begin{eqnarray}
A_{\CP}(f;t)\equiv\frac{N[\Bbar^0(t)\to f]-N[B^0(t)\to f]}
{N[\Bbar^0(t)\to f]+N[B^0(t)\to f]}=S_f\sin(\Delta m\,t)+
A_f\cos(\Delta m\,t),
\end{eqnarray}
where $N[\Bbar^0/B^0(t)\to f]$ is the number of $\Bbar^0/B^0$s that
decay to a common $\CP$ eigenstate $f$ after time $t$ and $\Delta m$
is the mass difference between the two neutral $B$ mass eigenstates.
The coefficients $S_f=\frac{2\,Im(\lambda_f)}{|\lambda_f|^2+1}$
and $A_f=\frac{|\lambda_f|^2-1}{|\lambda_f|^2+1}$ are functions
of the ratio of the decay amplitudes with and without mixing,
$\lambda_f=\frac{q}{p}\Abar_f/{\cal A}_f$. Here the ratio
$\frac{q}{p}$ accounts for $B^0\Bbar^0$ mixing; $S_f$ and
$A_f$ are the measure of mixing-induced and direct $\CP$ violation,
respectively. Note that BaBar uses a notation $C_f=-A_f$.

In the scenario where the neutral $B$ meson decays involve only
$b\to u$ transition, $\lambda_f$ is $e^{-i2(\phi_1+\phi_3)}=
e^{i2\phi_2}$ resulting in $S_f=\sin(2\phi_2)$ and $A_f=0$. This
is expected as direct $\CP$ violation, i.e., nonzero $A_f$
arises when there are at least two competing decay amplitudes of
different weak phase. Had that been the case, a time-dependent
$\CP$ analysis in the mode would have provided a clean access to
$\phi_2$. The real situation, however, is murkier due to the presence
of possible $b\to d$ penguin (loop) diagram, which does not carry
the same weak phase as the tree amplitude. As a result, the measured
$A_f$ is no longer equal to zero and the mixing-induced term becomes
$S_f=\sqrt{1-A^2_f}\sin(2\phi^{\rm eff}_2)$, where $\phi^{\rm eff}_2$
is an effective $\phi_2$ that differs from the true value because of
the penguin pollution.

\begin{figure}[!hbtp]
\centering
\includegraphics[width=0.5\textwidth]{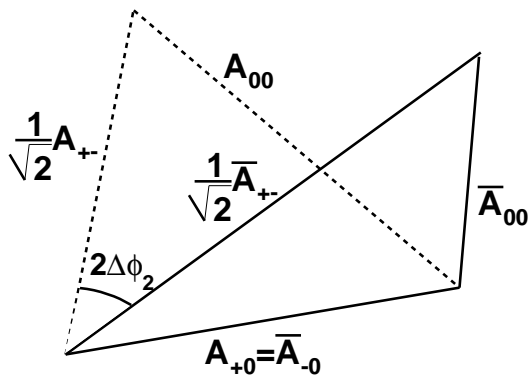}
\caption{Graphical representation of the isospin relations.}
\label{fig:su2}
\end{figure}

Gronau and London have proposed a model-independent method~\cite{ref:su2}
based on SU(2) isospin for extracting the difference $\Delta\phi_2=
\phi_2-\phi^{\rm eff}_2$, which would allow one to recover the
true value of $\phi_2$. Let us denote the $B^{ij}\to h^ih^j$ and
$\Bbar^{ij}\to h^ih^j$ ($h=\pi$ or $\rho$ and $i,j=+,-,0$) decay
amplitudes ${\cal A}_{ij}$ and $\Abar_{ij}$, respectively. Assuming
SU(2) isospin symmetry, these amplitudes are related by
\begin{eqnarray}
\frac{1}{\sqrt{2}}{\cal A}_{+-}+{\cal A}_{00}={\cal A}_{+0},\,\,\,\,
\frac{1}{\sqrt{2}}\Abar_{+-}+\Abar_{00}=\Abar_{-0},
\end{eqnarray}
which can be represented graphically as triangles in the complex plane
(see Fig.~\ref{fig:su2}). Neglecting electroweak penguins, $|{\cal A}_{+0}|=
|\Abar_{-0}|$ (evidence for large $\CP$ asymmetry in $B^{\pm}\to h^{\pm}h^0$,
if found, would show that such contributions cannot be neglected and
hence would invalidate the equality). If the global phase of all
${\cal A}_{ij}$ is chosen such that ${\cal A}_{+0}=\Abar_{-0}$, then
the phase difference between ${\cal A}_{+-}$ and $\Abar_{+-}$ would
be $2\Delta\phi_2$, as shown in Fig.~\ref{fig:su2}. Since both ${\cal A}$
and $\Abar$ triangles have two possible orientations, up and down, the
isospin method carries a four-fold ambiguity, which comes on top of
the two-fold ambiguity arising from the fact that only the sine of
$\phi^{\rm eff}_2$ is measured in time-dependent analyses of $B^0\to
h^+h^-$.

\section{{\boldmath $\phi_2$} from {\boldmath $B\to\pi\pi$}}

In order to determine $\phi_2$ in $B\to\pi\pi$ decays using the isospin
method, one needs six observables: mixing-induced as well as direct $\CP$
asymmetries in $B^0\to\pi^+\pi^-$, branching fractions for $B^0\to\pi^+\pi^-$,
$B^+\to\pi^+\pi^0$, and $B^0\to\pi^0\pi^0$ and the direct $\CP$ asymmetry in
$B^0\to\pi^0\pi^0$. Table~\ref{tab:pipi} summarizes results on these observables
as measured by Belle~\cite{ref:pipi-belle} and BaBar~\cite{ref:pipi-babar}.
Except for $A_{\pi^0\pi^0}$ both experiments exhibit similar sensitivities.
In this table, we also present $\CP$ asymmetry results for the channel
$B^+\to\pi^+\pi^0$, which are found to be consistent with zero, corroborating
our earlier assumption of negligible electroweak penguin contribution.

\begin{figure}[!hbtp]
\centering
\includegraphics[height=0.28\textheight]{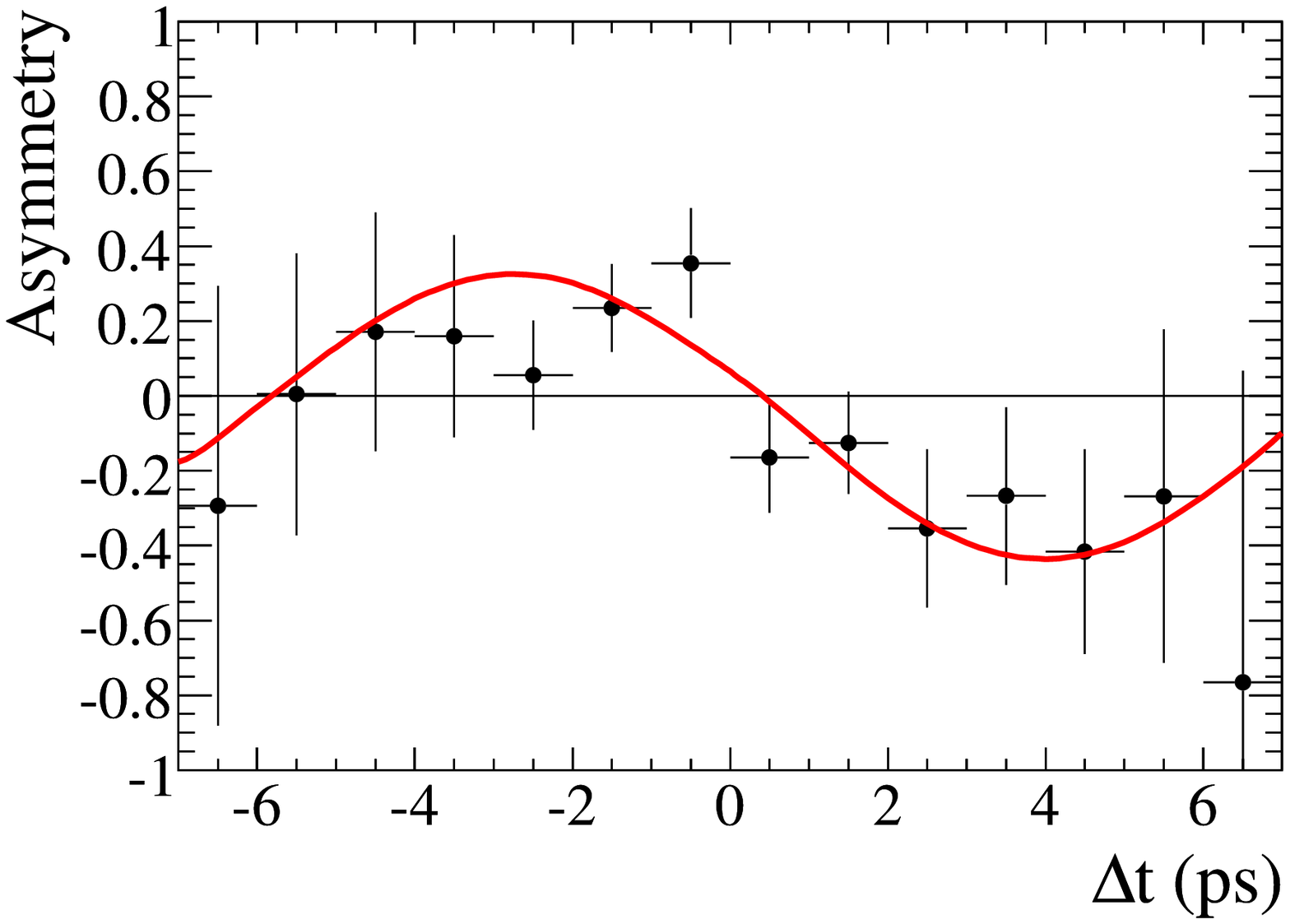}
\includegraphics[height=0.28\textheight]{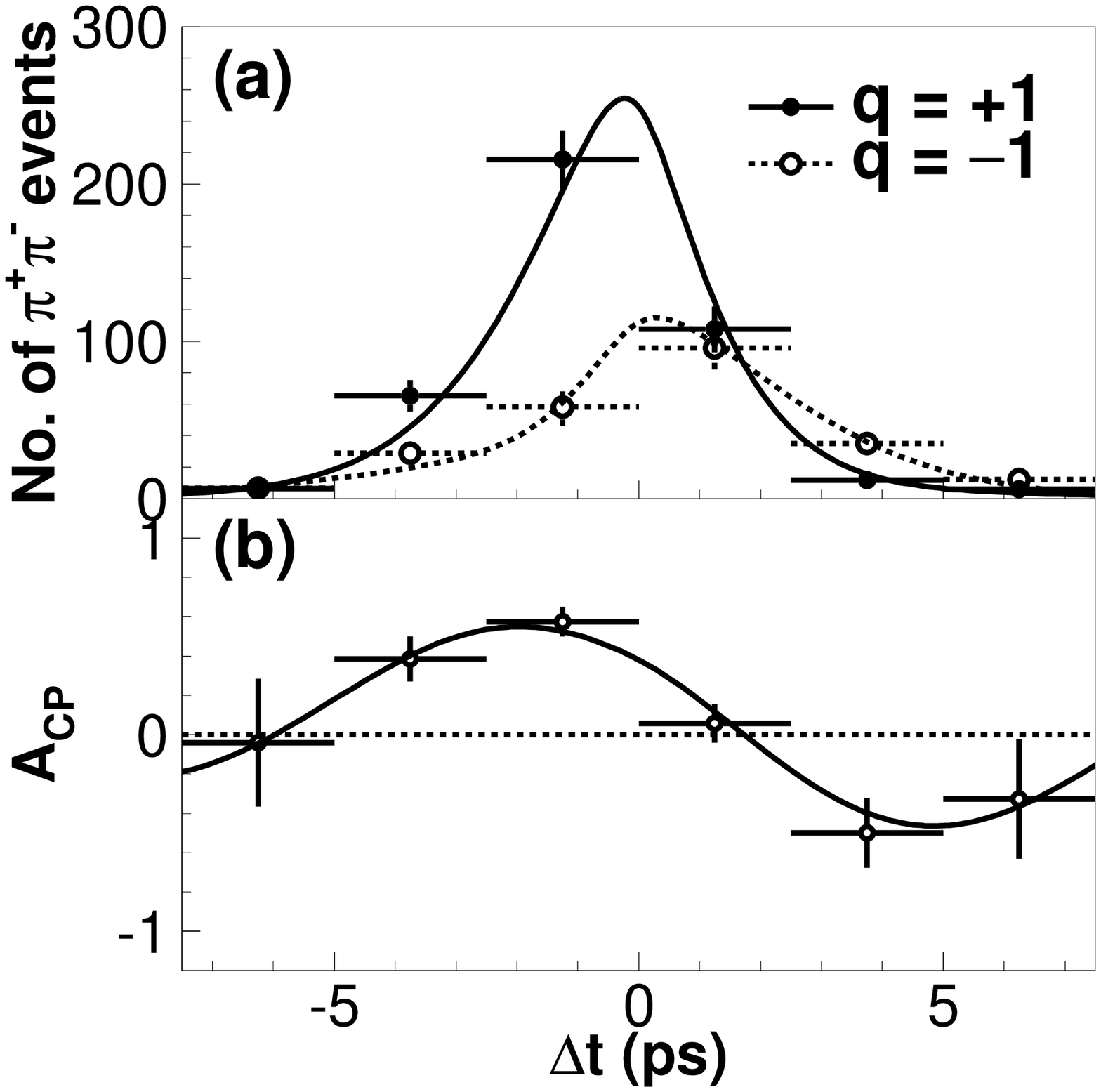}
\caption{The left plot shows the time-dependent asymmetry measured by
BaBar in $B^0\to\pi^+\pi^-$. The right plots from Belle are (a) results
of the fit to proper time distributions for the aforementioned decays
tagged as either $B^0({\rm q}=+1)$ or $\Bbar^0({\rm q}=-1)$, and (b)
the resultant asymmetry. Mixing-induced $\CP$ violation can be clearly
seen from the asymmetry plots, and the height difference between $B^0$-
and $\Bbar^0$-tagged decays (see the top right plot) is a signature of
direct $\CP$ violation.} 
\label{fig:cp-pipi}
\end{figure}

In Fig.~\ref{fig:cp-pipi} we plot results of the time-dependent $\CP$ fit
performed by Belle and BaBar in the decays $B^0\to\pi^+\pi^-$. Both
experiments observe nonzero mixing-induced $\CP$ asymmetry. As far as the
direct $\CP$ violation part is concerned Belle observe it with a significance
of $5.5$ standard deviations ($\sigma$), whereas BaBar find a $3.0\sigma$
evidence. Using the isospin analysis, Belle determine four different solutions
consistent with their data. The solution consistent with the SM inferred
value, $\left(100\,^{+\,5}_{-\,7}\right)^\circ$~\cite{ref:ckmfitter}, yields
$\phi_2=(97\pm11)^\circ$. Belle also exclude the interval $11^\circ<\phi_2
<79^\circ$ at the $95\%$ confidence level (CL). BaBar on the other hand set
an $1\sigma$ range for $\phi_2$ between $71^\circ$ and $109^\circ$,
excluding $\phi_2\in[23^\circ,67^\circ]$ at the $90\%$ CL. As evident from
Table~\ref{tab:pipi} the current precision on the measurements of $B^0\to
\pi^0\pi^0$, related to ${\cal A}_{00}$ and $\Abar_{00}$ of the isospin
relations in Eq.~(2), is the limiting factor in the extraction of $\phi_2$
in $B\to\pi\pi$.

\begin{table}[!hbtp]
\caption{Summary of physics observables measured in $B\to\pi\pi$ decays.
First uncertainties are statistical and second are systematic. Values in
square brackets denote numbers of $B\Bbar$ events (M stands for million)
in the data sample used in the analysis.}
\resizebox{\textwidth}{!}{
\begin{tabular}{llclc}  
\hline\hline
             &\multicolumn{1}{c}{Belle}&&\multicolumn{1}{c}{BaBar}&\\ \hline
$S_{\pi^+\pi^-}$&$-0.61\pm0.10\pm0.04$&[$535$ M]&$-0.68\pm0.10\pm0.03$&[$467$ M]\\
$A_{\pi^+\pi^-}$&$+0.55\pm0.08\pm0.05$&[$535$ M]&$+0.25\pm0.08\pm0.02$&[$467$ M]\\
${\cal B}(B^0\to\pi^+\pi^-)$&$(5.1\pm0.2\pm0.2)\times10^{-6}$&[$449$ M]&$(5.5\pm0.4\pm0.3)\times10^{-6}$&[$227$ M]\\
${\cal B}(B^+\to\pi^+\pi^0)$&$(6.5\pm0.4\pm0.4)\times10^{-6}$&[$449$ M]&$(5.02\pm0.46\pm0.29)\times10^{-6}$&[$383$ M]\\
$A_{\pi^+\pi^0}$&$+0.07\pm0.06\pm0.01$&[$535$ M]&$+0.03\pm0.08\pm0.01$&[$383$ M]\\
${\cal B}(B^0\to\pi^0\pi^0)$&$(1.1\pm0.3\pm0.1)\times10^{-6}$&[$535$ M]&$(1.83\pm0.21\pm0.13)\times10^{-6}$&[$467$ M]\\
$A_{\pi^0\pi^0}$&$+0.44\,^{+\,0.73}_{-\,0.62}\,^{+\,0.04}_{-\,0.06}$&[$535$ M]&$+0.43\pm0.26\pm0.05$&[$467$ M]\\
\hline\hline
\end{tabular}}
\label{tab:pipi}
\end{table}

\section{{\boldmath $\phi_2$} from {\boldmath $B\to(\rho\pi)^0$}}

The decays $B^0\to\rho^+\pi^-$, $B^0\to\rho^-\pi^+$, and $B^0\to\rho^0\pi^0$
[collectively referred to as $B\to(\rho\pi)^0$] are not a $\CP$ eigenstate
unlike $B\to\pi\pi$ or $B\to\rho\rho$. As a result, the time-dependent
parameters $S_f$ and $A_f$ measured in these decays are no more purely $\CP$
violating in nature; they contain a $\CP$ conserving part too. The problem
gets further complicated as the decays involve four isospin amplitudes
leading to $12$ unknowns in the isospin pentagon. Synder and Quinn have
suggested a theoretically cute idea~\cite{ref:synder-quinn} to determine
$\phi_2$ without any discrete, trigonometric ambiguity using a time-dependent
Dalitz plot (TDPA) analysis of $\pi^+\pi^-\pi^0$, which is the end product
of the $B\to(\rho\pi)^0$ decays. The basic philosophy behind their idea is
to exploit the strong phase variation of the interfering $\rho$ resonances
over the three-pion Dalitz plot.

\begin{figure}[!htb]
\centering
\includegraphics[height=0.25\textheight]{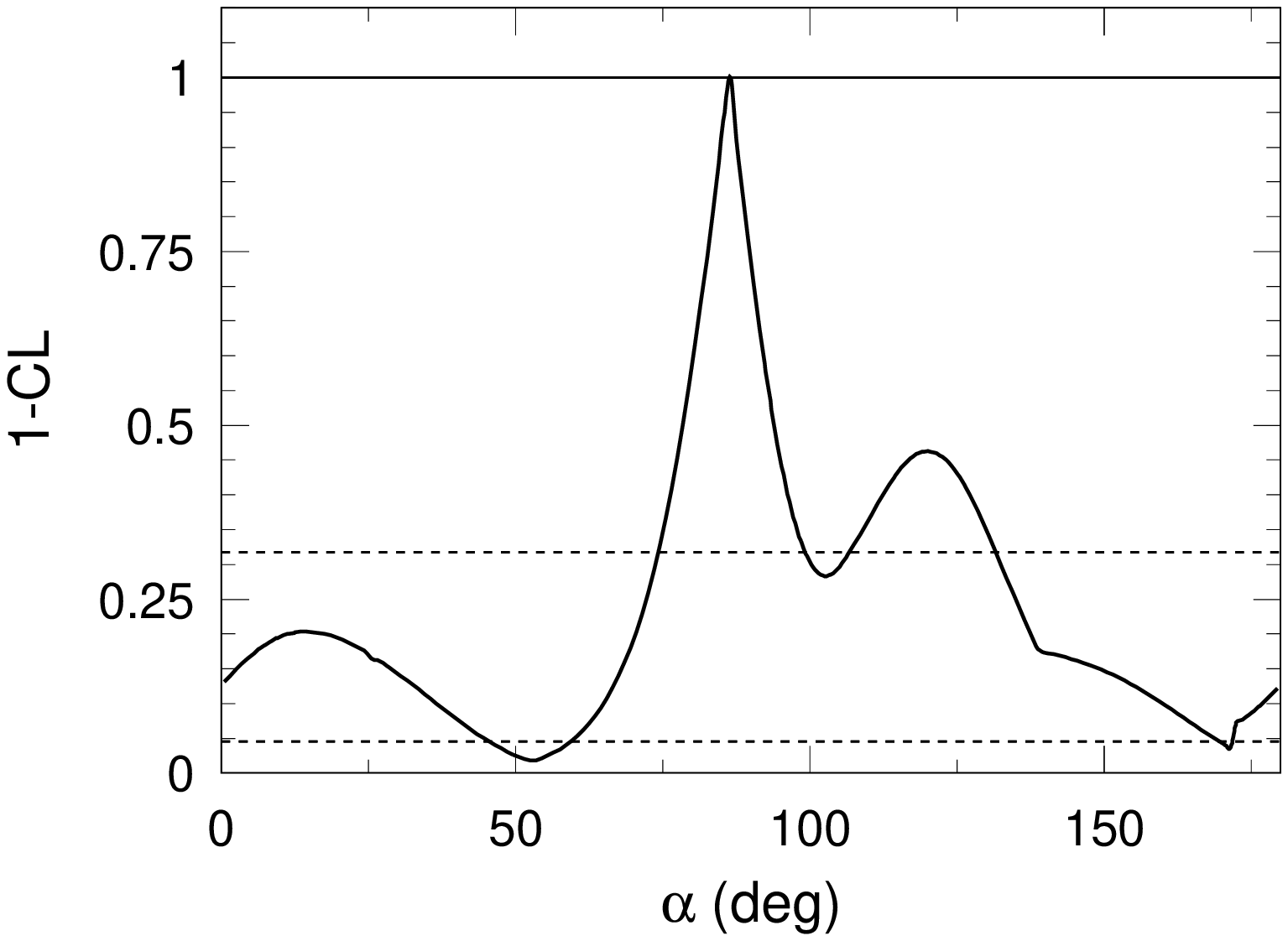}
\includegraphics[height=0.25\textheight]{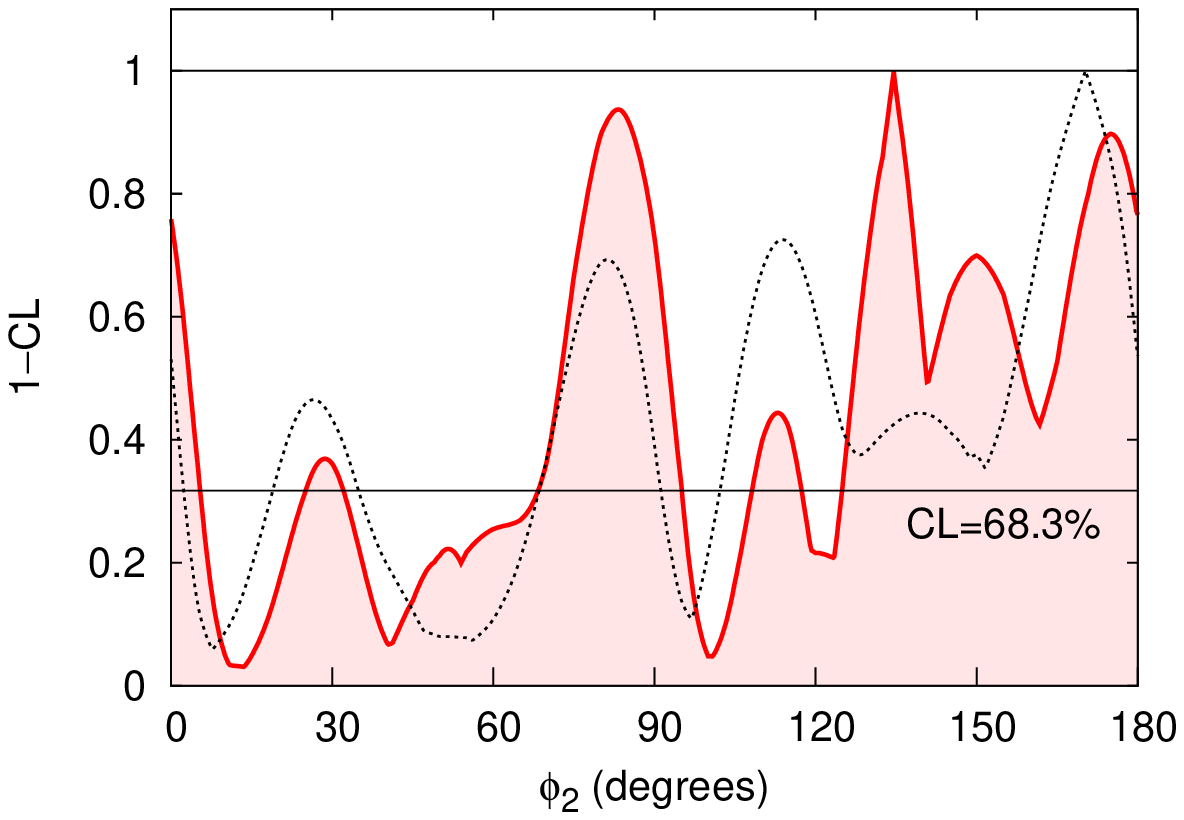}
\caption{$1-$CL versus $\phi_2$ from BaBar (left) and Belle (right)
 obtained in the analysis of $B\to(\rho\pi)^0\to\pi^+\pi^-\pi^0$.
 In the left plot the dashed horizontal lines near $1-$CL $= 0.317$
 and $0.050$ represent the $1\sigma$ and $2\sigma$ confidence intervals,
 respectively. For Belle, the dashed curve corresponds to the
 nine-parameter fit (including $\phi_2$) while the solid (red) curve
 includes additional constraints from $B^+\to\rho^0\pi^+$ and
 $B^+\to\rho^+\pi^0$.}
\label{fig:phi2-rhopi}
\end{figure}

Both Belle~\cite{ref:rhopi-belle} and BaBar~\cite{ref:rhopi-babar} have
performed the TDPA analysis using a dataset of $449$\,M and $375$\,M $B\Bbar$
events, respectively. These analyses are essentially based on a fit to $27$
bilinear coefficients~\cite{ref:quinn-silva}. All the $27$ observables are
not independent. There are in fact $12$ free parameters in the fit corresponding
to the six amplitudes ${\cal A}_i$ and $\Abar_i$ ($i=+,-,$ and $0$ represent
$\rho^+\pi^-$, $\rho^- \pi^+$, and $\rho^0\pi^0$, respectively). After factoring
out the overall normalization and phase factor along with two isospin relations
for the neutral $B$ decays, one is left with eight free parameters. To
constrain $\phi_2$, a $\chi^2$ is built out of the $26$ bilinear
coefficients (one of them is fixed as the overall normalization) after
taking correlations between them into account. A scan is performed over
all possible values of $\phi_2$, where each of the eight independent
parameters are varied in order to minimize the $\chi^2$. The resulting
change in the $\chi^2$ from its minimum is translated into a CL. In Fig.
\ref{fig:phi2-rhopi} we present results of $1-$CL versus $\phi_2$ obtained
by the two experiments. BaBar quote $\phi_2=\left(87\,^{+\,45}_{-\,13}\right)^\circ$;
almost no constraint is achieved at the $2\sigma$ interval. In conjunction
with additional constraints from the charged modes $B^+\to\rho^0\pi^+$ and
$B^+\to\rho^+\pi^0$~\cite{ref:hfag}, Belle derive the constraint $68^\circ
<\phi_2<95^\circ$ at $68.3\%$ CL for the solution consistent with the
SM~\cite{ref:ckmfitter}.

\section{{\boldmath $\phi_2$} from {\boldmath $B\to\rho\rho$}}

The decays $B\to\rho\rho$ comes with a complication that two fairly wide,
spin-1 mesons are present in the final state having a relative orbital
angular momentum, $L=0,1,2$. As the $\CP$ eigenvalue for $B^0\to\rho^+
\rho^-$ is $(-1)^L$, it becomes necessary to separate out the two distinct
$\CP$ components ($\CP$-even for $L=0,2$ and -odd for $L=1$) through an
angular analysis for constraining $\phi_2$. In other word, the extraction
of $\phi_2$ requires knowledge of polarization. At the $B$ factories, one
simultaneously measures the branching fraction as well as the fraction of
longitudinal polarization ($f_L$) by fitting the angular decay rate
\begin{eqnarray}
\frac{d^2N}{d\cos\theta_1d\cos\theta_2}=4f_L\cos^2\theta_1\cos^2\theta_2
+(1-f_L)\sin^2\theta_1\sin^2\theta_2,
\end{eqnarray}
where $\theta_1(\theta_2)$ is the helicity angle of the $\rho^+(\rho^-)$
meson. The two terms in the right hand side correspond to the contribution
of the longitudinal and transverse component, respectively.
Table~\ref{tab:rhorho} summarizes results from Belle~\cite{ref:rhorho-belle}
and BaBar~\cite{ref:rhorho-babar} on the branching fraction, $f_L$, and
time-dependent $\CP$ violation parameters $S_{\rho^+\rho^-}$ and $A_{\rho^+
\rho^-}$. The $f_L$ value being very much closer to unity tells us that
$B^0\to\rho^+\rho^-$ is dominated by the $\CP$-even, longitudinal amplitude
(the same conclusion also holds for $B^+\to\rho^+\rho^0$ and $B^0\to\rho^0
\rho^0$). Therefore, the transverse component can be safely ignored in the
extraction $\phi_2$. The measured values $S_{\rho^+\rho^-}$ and $A_{\rho^+
\rho^-}$ are found to be consistent with zero, ruling out a large contribution
from the $b\to d$ penguin amplitude.

\begin{table}[!hbtp]
\caption{Summary of physics observables measured in $B\to\rho\rho$ decays.
First uncertainties are statistical and second are systematic. Values in
square brackets denote numbers of $B\Bbar$ events in the data sample used
in the analysis.}
\resizebox{\textwidth}{!}{
\begin{tabular}{llclc}  
\hline\hline
             &\multicolumn{1}{c}{Belle}&&\multicolumn{1}{c}{BaBar}&\\ \hline
${\cal B}(B^0\to\rho^+\rho^-)$&$(22.8\pm3.8\,^{+\,2.3}_{-\,2.6})\times10^{-6}$&[275 M]&$(25.5\pm2.1\,^{+\,3.6}_{-\,3.9})\times10^{-6}$&[384 M]\\
$f_L(B^0\to\rho^+\rho^-)$&$0.941\,^{+\,0.034}_{-\,0.040}\pm0.030$&[275 M]&$0.992\pm0.024\,^{+\,0.026}_{-\,0.013}$&[384 M]\\
$S_{\rho^+\rho^-}$&$+0.19\pm0.30\pm0.08$&[535 M]&$-0.17\pm0.20\,^{+\,0.05}_{-\,0.06}$&[384 M]\\
$A_{\rho^+\rho^-}$&$+0.16\pm0.21\pm0.08$&[535 M]&$-0.01\pm0.15\pm0.06$&[384 M]\\
${\cal B}(B^+\to\rho^+\rho^0)$&$(31.7\pm7.1\,^{+\,3.8}_{-\,6.7})\times10^{-6}$&[85 M]&$(23.7\pm1.4\pm1.4)\times10^{-6}$&[465 M]\\
$f_L(B^+\to\rho^+\rho^0)$&$0.95\pm0.11\pm0.02$&[85 M]&$0.950\pm0.015\pm0.006$&[465 M]\\
$A_{\rho^+\rho^0}$&$+0.00\pm0.22\pm0.03$&[85 M]&$-0.054\pm0.055\pm0.010$&[465 M]\\
${\cal B}(B^0\to\rho^0\rho^0)$&$<1.0\times10^{-6}$ at $90\%$ CL&[657 M]&$(0.92\pm0.32\pm0.14)\times10^{-6}$&[465 M]\\
$f_L(B^0\to\rho^0\rho^0)$&\multicolumn{1}{c}{--}&--&$0.75\,^{+\,0.11}_{-\,0.14}\pm0.04$&[465 M]\\
$S_{\rho^0\rho^0}$&\multicolumn{1}{c}{--}&--&$+0.3\pm0.7\pm0.2$&[465 M]\\
$A_{\rho^0\rho^0}$&\multicolumn{1}{c}{--}&--&$-0.2\pm0.8\pm0.3$&[465 M]\\
\hline\hline
\end{tabular}}
\label{tab:rhorho}
\end{table}

\begin{figure}[!htb]
\centering
\includegraphics[height=0.22\textheight]{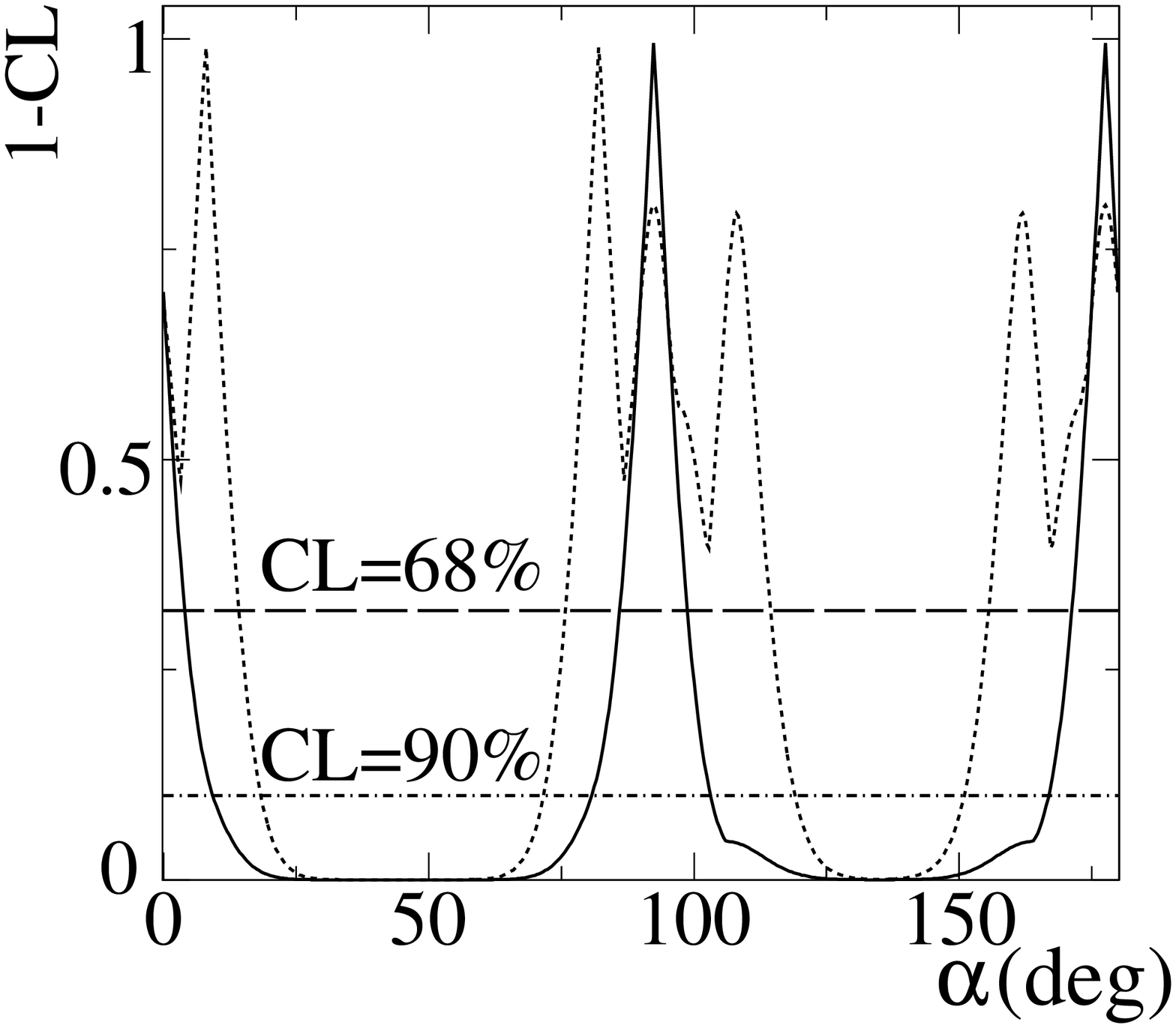}
\includegraphics[height=0.22\textheight]{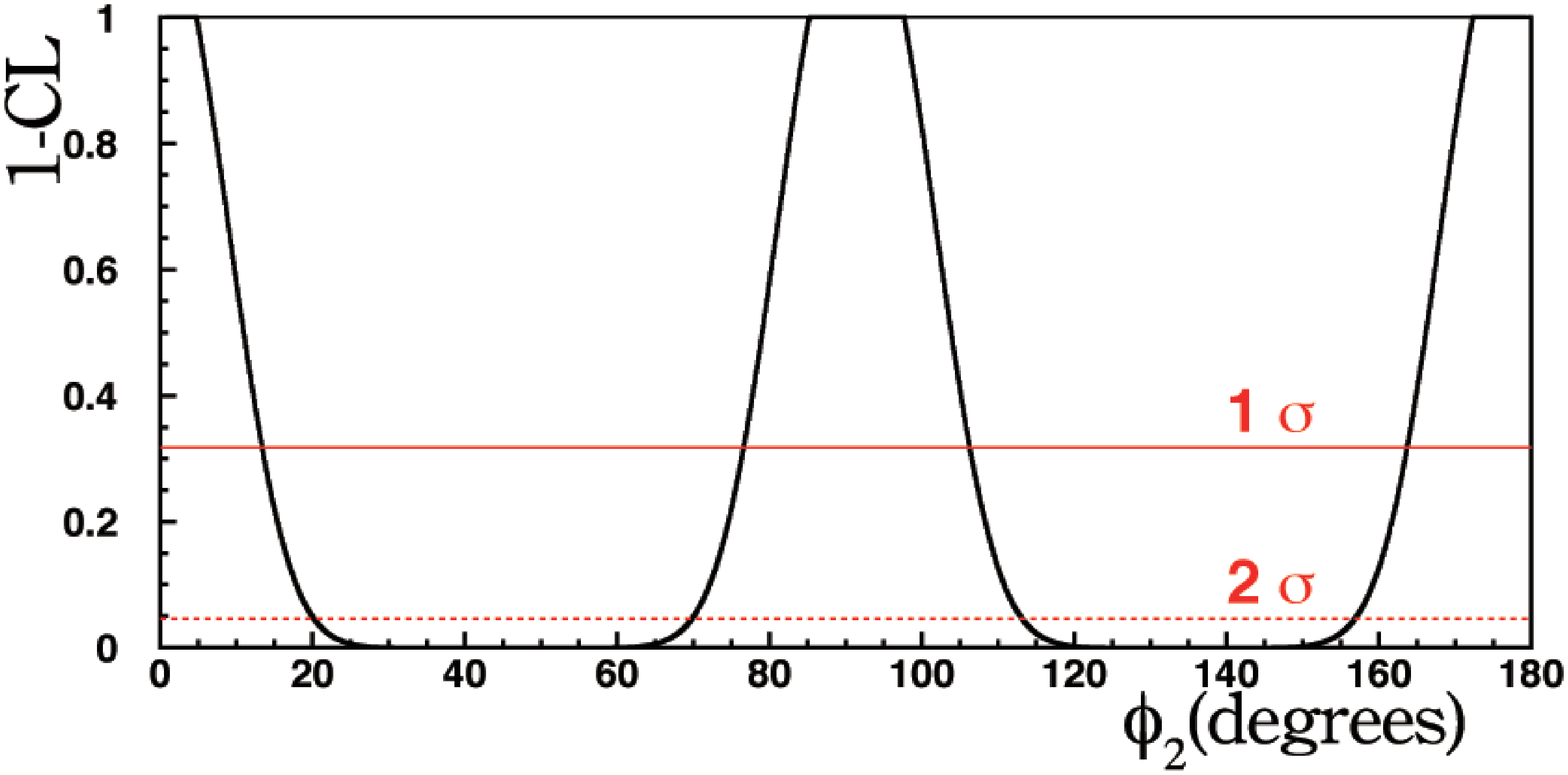}
\caption{$1-$CL versus $\phi_2$ from BaBar (left) and Belle (right)
 obtained with the analysis of $B\to\rho\rho$. BaBar's constraint
 is based solely on their results. The dashed curve indicates situation
 prior to the updated result of $B^+\to\rho^+\rho^0$ while the solid
 curve includes that. Belle use their own results on $B^0\to\rho^0
 \rho^0$ plus world-averages for the rest. The latter exercise was
 carried out before BaBar's update on $B^+\to\rho^+\rho^0$, and a
 plateau is present as there is no constraint on $A_{\rho^0\rho^0}$.}
\label{fig:phi2-rhorho}
\end{figure}

Results from Belle and BaBar on $B^+\to\rho^+\rho^0$ and $B^0\to\rho^0
\rho^0$ are also presented in Table~\ref{tab:rhorho}. In case of
$B^+\to\rho^+\rho^0$ BaBar obtain a pretty large branching fraction,
which leads to a strong constraint on the isospin triangles and
subsequently to a precise determination of $\phi_2$ (see the discussion
below), and a $\CP$ asymmetry consistent with zero, showing no evidence
for isospin violation. Belle's earlier results based on a much smaller
dataset than available today agree with these measurements -- obviously
with larger errors. For $B^0\to \rho^0\rho^0$, although it is presumed
to be easy to identify having four charged pions in the final state,
one suffers due to relatively low branching fraction in the face of
multiple backgrounds with the identical final state. BaBar report a
first evidence for the decay with a $3.1\sigma$ significance in
addition to performing a time-dependent study. Belle on the other
hand quote a $90\%$ CL upper limit on the branching fraction. In
Fig.~\ref{fig:phi2-rhorho} we show the constraints on $\phi_2$
resulting from $B\to\rho\rho$ in. The results are $\phi_2=(91.7\pm14.9)^\circ$
and $\left( 92.4\,^{+\,6.0}_{-\,6.5}\right)^\circ$ from Belle and BaBar,
respectively.

\section{Closing Words on {\boldmath $\phi_2$}}

The current world-average of $\phi_2$~\cite{ref:ckmfitter} including all
relevant measurements of $B\to\pi\pi$, $B\to\rho\pi$ and $B\to\rho\rho$
from Belle and BaBar reads $\left(89.0\,^{+\,4.4}_{-\,4.2}\right)^\circ$ --
almost a precision measurement. Another averaging method~\cite{ref:utfit}
based on a different statistical treatment also yields a compatible result.
The world-average value is mostly decided by the results of $B\to\rho\rho$.
In particular, the large branching fraction for $B^+\to\rho^+\rho^0$ relative
to $B^0\to\rho^0\rho^0$ reported by BaBar is responsible in collapsing two
previously degenerated mirror solutions of the isospin relations, inherent
to the method, into a single one. The two isospin triangles in fact do not
close any more~\cite{ref:ckmfitter}. This may sound a bit of jocular; to start with we assume
the SU(2) constraint for extracting $\phi_2$, while the end result is a
violation of the triangular relation, but yields a precision measurement!
We are truly in a lucky situation, and therefore it is very important to
have the updated results from Belle with their full $\FourS$ data sample.
Furthermore, more precise results on $B^0\to\pi^0\pi^0$, which can only
be carried out at the pristine $e^+e^-$ environment of the $B$ factories,
would play a good supporting role. Last but not least, one should not
forget about the mode $B\to(\rho\pi)^0$, which provides a clean measure
of $\phi_2$ without any trigonometric ambiguity. It would be imperative
to carry out these important measurements with much more data at the
next-generation flavor factories, SuperKEKB in Japan~\cite{ref:superkekb}
and SuperB in Italy~\cite{ref:superb}.

\section{The {\boldmath $K\pi$} Puzzle}

As alluded earlier, direct $\CP$ violation arises when we have contributions
from two competing amplitudes of different weak phases. For the decays
$B^0\to K^+\pi^-$ and $B^+\to K^+\pi^0$, that are assumed to proceed via
similar Feynman's diagrams at the tree level, one expects to see the
magnitude of $\CP$ violation to be same. In contrast, the measured $\CP$
asymmetry in $B^0\to K^+\pi^-$ is $(-9.8\,^{+\,1.2}_{-\,1.1})\%$, whereas
the same for $B^+\to K^+\pi^0$ reads $(+5.0\pm 2.5)\%$. This apparent
large gap~\cite{ref:hfag} between the two measurements [$\Delta A_{K\pi}
=(-14.4\pm2.9)\%$] is better known as the $K\pi$ puzzle. The jury is out
whether this could be due to a large contribution from the color-suppressed
tree diagram, or an enhanced electroweak penguin amplitude, or Pauli
blocking, as it has been recently claimed~\cite{ref:kpi-lipkin}. But still,
it is fair to say that possible smoking gun for new physics~\cite{ref:np-kpi}
is not unequivocally ruled out. On this account, it would be good to improve
the precision on $\CP$ violation results for $B^0\to K^0\pi^0$ in order to
test the sum rule~\cite{ref:sumrule}. Data from the super flavor factories
\cite{ref:superkekb,ref:superb} are expected to play a decisive role on
resolving this $K\pi$ conundrum.


\Acknowledgements
We thank Karim Trabelsi for reading the proceedings and making useful
suggestions. This work is supported in parts by the Department of Atomic
Energy and the Department of Science and Technology of India.


\begin{thebibliography}{99}

\bibitem{ref:ckm}
 N. Cabibbo, Phys.\ Rev.\ Lett.\ {\bf 10}, 531 (1963);
 M. Kobayashi and T. Maskawa, Prog.\ Theor.\ Phys.\ {\bf 49}, 652 (1973).

\bibitem{ref:belle}
 A. Abashian {\it et al.} (Belle Collaboration), Nucl.\ Instrum.\ Methods
 Phys.\ Res., Sect. A {\bf 479}, 117 (2002).

\bibitem{ref:babar}
 B. Aubert {\it et al.} (BaBar Collaboration), Nucl.\ Instrum.\ Methods
 Phys.\ Res., Sect. A {\bf 479}, 1 (2002).

\bibitem{ref:lhcb}
 A. Augusto Alves {\it et al.} (LHCb Collaboration), JINST {\bf 3},
 S08005 (2008).

\bibitem{ref:su2}
 M. Gronau and D. London, Phys.\ Rev.\ Lett.\ {\bf 65}, 3381 (1990).

\bibitem{ref:pipi-belle}
 H. Ishino {\it et al.} (Belle Collaboration), Phys.\ Rev.\ Lett.\ {\bf 98}, 211801 (2007);
 S.-W. Lin {\it et al.} (Belle Collaboration), Phys.\ Rev.\ Lett.\ {\bf 99}, 121601 (2007);
 S.-W. Lin {\it et al.} (Belle Collaboration), Nature {\bf 452}, 332 (2008);
 K. Abe {\it et al.} (Belle Collaboration), arXiv:hep-ex/0610065.

\bibitem{ref:pipi-babar}
 B. Aubert {\it et al.} (BaBar Collaboration), arXiv:0807.4226[hep-ex];
 B. Aubert {\it et al.} (BaBar Collaboration), Phys.\ Rev.\ D {\bf 75}, 012008 (2007);
 B. Aubert {\it et al.} (BaBar Collaboration), Phys.\ Rev.\ D {\bf 76}, 091102 (2007).

\bibitem{ref:ckmfitter}
 J. Charles {\it et al.} (CKMfitter Group), Eur.\ Phys.\ J.\ C {\bf 41}, 1 (2005);
 For online updates see {\tt http://ckmfitter.in2p3.fr/}.

\bibitem{ref:synder-quinn}
 A.E. Synder and H.R. Quinn, Phys.\ Rev.\ D {\bf 48}, 2139 (1993).

\bibitem{ref:rhopi-belle}
 A. Kusaka {\it et al.} (Belle Collaboration), Phys.\ Rev.\ Lett.\ {\bf 98}, 221602 (2007);
 A. Kusaka {\it et al.} (Belle Collaboration), Phys.\ Rev.\ D {\bf 77}, 072001 (2008).

\bibitem{ref:rhopi-babar}
 B. Aubert {\it et al.} (BaBar Collaboration), Phys.\ Rev.\ D {\bf 76}, 012004 (2007).

\bibitem{ref:quinn-silva}
 H.R. Quinn and J.P. Silva, Phys.\ Rev.\ D {\bf 62}, 054002 (2000).

\bibitem{ref:hfag}
 E. Barberio {\it et al.} (Heavy Flavor Averaging Group), arXiv:hep-ex/0603003;
 For online updates see {\tt http://www.slac.stanford.edu/xorg/hfag/}.

\bibitem{ref:rhorho-belle}
 A. Somov {\it et al.} (Belle Collaboration), Phys.\ Rev.\ D {\bf 76}, 011104 (2007);
 A. Somov {\it et al.} (Belle Collaboration), Phys.\ Rev.\ Lett.\ {\bf 96}, 171801 (2006);
 J. Zhang {\it et al.} (Belle Collaboration), Phys.\ Rev.\ Lett.\ {\bf 91}, 221801 (2003).

\bibitem{ref:rhorho-babar}
 B. Aubert {\it et al.} (BaBar Collaboration), Phys.\ Rev.\ D {\bf 76}, 052007 (2007);
 B. Aubert {\it et al.} (BaBar Collaboration), Phys.\ Rev.\ Lett.\ {\bf 97}, 261801 (2006);
 B. Aubert {\it et al.} (BaBar Collaboration), Phys.\ Rev.\ D {\bf 78}, 071104 (2008).

\bibitem{ref:utfit}
 M. Bona {\it et al.} (UTfit Collaboration), JHEP {\bf 0507}, 028 (2005);
 For online updates see {\tt http://www.utfit.org/}.

\bibitem{ref:superkekb}
 T. Abe {\it et al.} (Belle II Collaboration), arXiv:1011.0352[hep-ex].

\bibitem{ref:superb}
 M. Bona {\it et al.} (SuperB Collaboration), arXiv:0709.0451[hep-ex].

\bibitem{ref:kpi-lipkin}
 H.J. Lipkin, contribution to these proceedings and arXiv:1105.3443[hep-ph].

\bibitem{ref:np-kpi}
 W.-S. Hou, M. Nagashima, and A. Soddu, Phys.\ Rev.\ Lett.\ {\bf 95}, 141601 (2005);
 R. Fleischer, S. Recksiegel, and F. Schwab, Eur.\ Phys.\ J.\ C {\bf 51}, 55 (2007);
 L. Hofer, D. Scherer, and L. Vernazza, JHEP {\bf 1102}, 080 (2011).

\bibitem{ref:sumrule}
 M. Gronau, Phys.\ Lett.\ B {\bf 627}, 82 (2005).

\end{thebibliography}
\end{document}